\documentclass{article}%
\usepackage[top=3cm, bottom=3cm, left=3cm, right=3cm]{geometry}
\usepackage{amsmath}
\usepackage{amsfonts}
\usepackage{amssymb}
\usepackage{graphicx}%
\begin{document}

\title{A pseudoscalar glueball and charmed mesons in the extended Linear Sigma Model %
\thanks{Presented at the 3rd International Conference on New Frontiers in Physics, 28th July - 6th August 2014, Kolymbari, Crete, Greece.}}%

\author{Walaa I. Eshraim\\\emph{Institute for Theoretical Physics, }\\\emph{Johann Wolfgang Goethe University,}\\\emph{Max-von-Laue-Str. 1, D-60438 Frankfurt am Main, Germany}}
\maketitle

\begin{abstract}
In the framework of the so-called extended linear sigma model
(eLSM), we include a pseudoscalar glueball with a mass of 2.6 GeV
(as predicted by Lattice-QCD simulations) and we compute the two-
and three-body decays into scalar and pseudoscalar mesons. This
study is relevant for the future PANDA experiment at the FAIR
facility. As a second step, we extend the eLSM by including the
charm quark according to the global $U(4)_R \times U(4)_L$ chiral
symmetry. We compute the masses, weak decay constants and strong
decay widths of open charmed mesons. The precise description of
the decays of open charmed states is important for the CBM
experiment at FAIR.

\end{abstract}


%
 
\section{Introduction}
\label{intro}
\indent The fundamental interactions of quarks and gluons are described by quantum chromodynamics (QCD). The development of an effective hadronic Lagrangian plays an important role in the description of the masses and the interactions of low-lying hadron resonances \cite{Amsler}. To this end, we developed the so-called extended Linear Sigma Model (eLSM) \cite{nf2D} in which (pseudo)scalar and (axial-)vector $q\overline{q}$ mesons and additional scalar and pseudoscalar glueball fields are the basic degrees of freedom. The eLSM emulates the global symmetries of the QCD Lagrangian: the global chiral symmetry (which is exact in the chiral limit), the discrete C, P, and T symmetries,
and the classical dilatation (scale) symmetry. When working with colorless hadronic degrees of
freedom, the local color symmetry of QCD is automatically preserved. In QCD (and thus also in the
eLSM) the global chiral symmetry is explicitly broken by non-vanishing quark masses and quantum
effects \cite{Hooft}, and spontaneously by a non-vanishing expectation value of the quark condensate in vacuum \cite{Vafa}. The
dilatation symmetry is broken explicitly by the logarithmic term of the dilaton potential, by the mass terms, and by the
$U(1)_A$ anomaly.\\
\indent The investigation of the properties of bound states of gluons, so-called glueballs, is an important field of research in hadronic physics. The glueball spectrum has been predicted by Lattice QCD \cite{Morningstar}, where the third lightest glueball is a pseudoscalar state ($J^{PC}=0^{-+}$), denoted as $\widetilde{G}\equiv gg$, with a mass of about $2.6$ GeV, which is studied in the present work.\\
\indent Studying the properties of the third most massive of all quarks, the charm quark, is an active field of hadronic physics \cite{brambilla}. Thus we extend the eLSM (which has shown success in describing the phenomenology of the nonstrange-strange mesons \cite{nf2D, dick}) from the three quark-flavor case \cite{dick} to the four quark-flavor case \cite{WFDP, WP, WFP} with only three new unknown parameters related to the charm sector. Chiral symmetry is strongly explicitly broken by the current charm quark mass.\\
\indent In the present work, we compute the two- and three-body decays of the lightest pseudoscalar glueball within the eLSM in the case of three flavors.
 Moreover, we calculate all meson masses in the eLSM including light and charmed
mesons, the (OZI-dominant) strong decays of open charmed mesons, and the weak decay constants of charmed mesons.\\
\indent These proceedings are organized as follows: in Sec. 2 we present the $U(N_f)_R \times U(N_f)_L$ extended Linear Sigma Model with (axial-)vector mesons. In Sec. 3 we present the vacuum properties of a pseudoscalar glueball within the eLSM. In Sec. 4 we outline the extension of the eLSM to $N_f=4$ and present the results for the masses, weak decay constants, and decay widths, and in Sec. 5 we  provide our conclusions and an outlook. 

\section{The extended Linear Sigma Model}\label{sec-1}

\indent In this section we present the eLSM Lagrangian which is constructed from two requirements stemming from the
underlying QCD theory: (i)
global chiral symmetry $U(N_f)_R \times U(N_f)_L$; (ii) dilatation
invariance, with the exceptions of the scale anomaly and the $U(1)_A$
anomaly. It is also invariant under the discrete symmetries $C$
and $P$. The Lagrangian of the eLSM for a generic number of
flavors $N_f$ \cite{nf2D, dick, WFDP, WP, WFP, Giacosa} takes the following form
\begin{align}
\mathcal{L}  & =\frac{1}{2}(\partial _{\mu
}G)^{2}-V_{dil}(G)+\mathrm{Tr}[(D^{\mu}\Phi)^{\dagger}(D^{\mu
}\Phi)]-m_{0}^{2}\left(  \frac{G}{G_{0}}\right)
^{2}\mathrm{Tr}(\Phi
^{\dagger}\Phi)-\lambda_{1}[\mathrm{Tr}(\Phi^{\dagger}\Phi)]^{2}\nonumber\\
& -\lambda
_{2}\mathrm{Tr}(\Phi^{\dagger}\Phi)^{2}+\mathrm{Tr}[H(\Phi+\Phi^{\dagger
})]
 +\mathrm{Tr}\left\{  \left[  \left(  \frac{G}{G_{0}}\right)  ^{2}%
\frac{m_{1}^{2}}{2}+\Delta\right]  \left[
(L^{\mu})^{2}+(R^{\mu})^{2}\right]
\right\} \nonumber\\
&  -\frac{1}{4}\mathrm{Tr}[(L^{\mu\nu})^{2}+(R^{\mu\nu})^{2}%
]-2\,\mathrm{Tr}[E\,\Phi^{\dagger}\Phi]
+c(\mathrm{det}\Phi
-\mathrm{det}\Phi^{\dagger})^{2}+ic_{\tilde{G}\Phi}\tilde{G}(\mathrm{det}\Phi
-\mathrm{det}\Phi^{\dagger})\nonumber\\
&  +i\frac{g_{2}}{2}\{\mathrm{Tr}(L_{\mu\nu}[L^{\mu},L^{\nu}])+\mathrm{Tr}%
(R_{\mu\nu}[R^{\mu},R^{\nu}])\}+\frac{h_{1}}{2}\mathrm{Tr}(\Phi^{\dagger}%
\Phi)\mathrm{Tr}[(L^{\mu})^{2}+(R^{\mu})^{2}]\nonumber\\
&+h_{2}\mathrm{Tr}[(\Phi R^{\mu })^{2}+(L^{\mu}\Phi)^{2}]
+2h_{3}\mathrm{Tr}(\Phi R_{\mu}\Phi^{\dagger}L^{\mu})+ \ldots\,,\,
\label{lag}%
\end{align}

where the field $G$ denotes the dilaton field and its potential \cite{Salomone}
reads
\begin{equation}
V_{dil}(G)=\frac{1}{4}\frac{m_{G}^2}{\Lambda_G^2}\bigg[G^4\rm{ln}\bigg(\frac{G}{\Lambda_G}\bigg)-\frac{G^4}{4}\bigg],
\end{equation}
in which the parameter $\Lambda_{G}\sim N_C\,\Lambda_{QCD}$ sets the
energy scale of the gauge theory. Moreover,
$D^{\mu}\Phi\equiv\partial^{\mu}\Phi-ig_{1}(L^{\mu}\Phi-\Phi
R^{\mu})$ is the covariant derivative and contains the
(pseudo)scalar quark-antiquark mesons,
$L^{\mu\nu}\equiv\partial^{\mu}L^{\nu}-\partial^{\nu}L^{\mu}$, and
$R^{\mu\nu }\equiv\partial^{\mu}R^{\nu}-\partial^{\nu}R^{\mu}$ are
the left-handed and right-handed field strength tensors which
describe the (axial-)vector mesons. In eLSM Lagrangian (\ref{lag}) the dots
refer to further chirally invariant terms listed in Ref.\
\cite{dick}: These terms do not affect the masses and decay widths
studied in the present work and we therefore omitted them. The terms involving the
matrices $H,\, E,$ and $\Delta$ are the mass terms and break the
dilatation and chiral symmetry due to nonzero current quark
masses. Note that the $E$ term is a new term added to the model to
account for the charm quark mass, whereas the term $c(\mathrm{det}\Phi
-\mathrm{det}\Phi^{\dagger})^{2}$ describes the $U(1)_A$ anomaly. The constant $c$ has dimension $[\rm{Energy}]^{4-2N_f}$. Furthermore, the
pseudoscalar field is the last field included in the model by the
$ic_{\tilde{G}\Phi}\tilde{G}(\mathrm{det}\Phi
-\mathrm{det}\Phi^{\dagger})$ term (details are provided in the next
section).\\
Spontaneous symmetry breaking takes place when $m_0^2<0$ and
the scalar-isoscalar fields condense as well as the
glueball field $G = G_0$, \cite{nf2D, dick, WFDP, WP, WFP}. To implement this breaking one has
to shift the scalar fields $G \rightarrow G_0+G$ and $\Phi\rightarrow
\rm{diag} \{\sqrt{2}\sigma_N, \,\sqrt{2}\sigma_N,...\}+\Phi$. Using the eLSM we study the vacuum properties of the lightest pseudoscalar
glueball $\tilde{G}$ in the case of $N_f=3$, and by extending it to $N_f=4$, we study the phenomenology of charmed mesons. Details will
be given in the following two sections.

\section{Phenomenology of a pseudoscalar glueball}\label{sec-2}

In this section we compute the decay widths of the pseudoscalar glueball
$\widetilde{G}$ into a scalar and a pseudoscalar meson,
$\widetilde{G}\rightarrow PS$ and into three pseudoscalar mesons
$\widetilde{G}\rightarrow PPP$.\\
In Ref.\cite{Eshraim} we considered an $N_f=3$ chiral Lagrangian which
describes the interaction between the pseudoscalar glueball
$\widetilde{G}\equiv|gg>$ and (pseudo-)scalar mesons as follows:

\begin{equation}
\mathcal{L}_{\tilde{G}}^{int}=ic_{\tilde{G}\Phi}\tilde{G}\left(
{\textrm{det}}\Phi-{\textrm{det}}\Phi^{\dag}\right)\label{intlag},
\end{equation}
where $c_{\tilde{G}\Phi}$ is a dimensionless coupling constant and
$\Phi$ reads for three flavours, $N_{f}=3$ \cite{dick}:

\begin{equation}
\Phi=(S^a+iP^a)\,t^a=\frac{1}{\sqrt{2}}\left(
\begin{array}
[c]{ccc}%
\frac{(\sigma_{N}+a_{0}^{0})+i(\eta_{N}+\pi^{0})}{\sqrt{2}} & a_{0}^{+}%
+i\pi^{+} & K_{S}^{+}+iK^{+}\\
a_{0}^{-}+i\pi^{-} & \frac{(\sigma_{N}-a_{0}^{0})+i(\eta_{N}-\pi^{0})}%
{\sqrt{2}} & K_{S}^{0}+iK^{0}\\
K_{S}^{-}+iK^{-} & \bar{K}_{S}^{0}+i\bar{K}^{0} & \sigma_{S}+i\eta_{S}%
\end{array}
\right)\,, \label{phimatex}%
\end{equation}
where $S$ and $P$ refer to scalar and pseudoscalar mesons,
respectively, while $t^a$ are generators of the group $U(N_f)$. As
discussed in Ref.\cite{Eshraim} the multiplet $\Phi$ transforms under
$U(3)_L\times U(3)_R$ chiral transformations as $\Phi \rightarrow
U_L \Phi U_R^\dagger$, whereas
$U_{L(R)}=e^{-i\theta_{L(R)}^{a}t^{a}}$ is an element of
$U(3)_{L(R)}$ and the pseudoscalar glueball $\widetilde{G}$ is
chirally invariant. We then conclude that the interaction
Lagrangian (\ref{intlag}) is invariant under $SU(3)_R \times
SU(3)_L$ and also invariant under parity ($\Phi \rightarrow
\Phi^\dagger$ and $\widetilde{G}\rightarrow-\widetilde{G}$), but not
invariant under the axial $U_A(1)$ transformation.\\
The states in the interaction Lagrangian are assigned as
physical resonances to light quark-antiquark states with mass
$\lesssim 2$ GeV \cite{Eshraim}. For the pseudoscalar sector $P$, the
fields $\overrightarrow{\pi}$ and $K$ represent the pion
isotriplet and the kaon isodoublet \cite{dick}. The bare fields $\eta_{N}%
\equiv|\bar{u}u+\bar{d}d\rangle/\sqrt{2}$ and
$\eta_{S}\equiv\bar{s}s\rangle$ are the nonstrange and strange contributions of the physical states $\eta$ and
$\eta^{\prime}$ which can be obtained by \cite{dick}:
\begin{equation}
\eta=\eta_{N}\cos\varphi+\eta_{S}\sin\varphi,\text{
}\eta^{\prime}=-\eta
_{N}\sin\varphi+\eta_{S}\cos\varphi, \label{mixetas}%
\end{equation}
where the mixing angle is $\varphi\simeq 44.6^{\circ}$ \cite{dick}. For
the scalar sector $S$, the field $\overrightarrow{a_0}$ is assigned
to the physical isotriplet state $a_0(1450)$ and the scalar kaon
fields $K_s$ to the resonance $K^\ast_0(1430)$. In the
scalar-isoscalar sector the bare nonstrange field
$\sigma_{N}\equiv\left\vert \bar{u}u+\bar{d}d\right\rangle
/\sqrt{2}$ corresponds to the resonance $f_0(1370)$ \cite{dick} but
the bare strange field $\sigma_{S}\equiv\left\vert \bar
{s}s\right\rangle$ can be assigned to the resonance $f_0(1710)$
\cite{dick} or $f_0(1500)$ \cite{Cheng}. \\
To compute the decays of the pseudoscalar glueball we have to
implement the spontaneous symmetry breaking by shifting the
scalar-isoscalar fields by their vacuum expectation values as
follows \cite{dick, Eshraim}:
\begin{equation}
\sigma_{N}\rightarrow\sigma_{N}+\phi_{N}\text{ and
}\sigma_{S}\rightarrow
\sigma_{S}+\phi_{S}\text{ ,} \label{shift}%
\end{equation}
where $\phi_N$ and $\phi_S$ represent the chiral-nonstrange and
strange condensates \cite{dick}, as the following
\begin{equation}
\phi_{N}=Z_{\pi}f_{\pi},\,\,\,\,\,\phi_{S}=\frac{2Z_{K}f_{K}-\phi_{N}%
}{\sqrt{2}}\;.
\end{equation}
Furthermore, we shift the axial-vector fields and redefine the
renormalization constant of the pseudoscalar fields to get rid of
the mixing between (axial-)vector and (pseudo)scalar states. \\
All parameters have been determined in Ref.\cite{dick}. Here, we present the theoretical results of the two-body decay $\widetilde{G} \rightarrow PS$ and the three-body decay $\widetilde{G} \rightarrow PPP$ \cite{Eshraim} as branching ratios,
where the mass of the pseudoscalar glueball $\widetilde{G}$ is
$2.6$ GeV predicted by Lattice QCD simulation \cite{Morningstar} which are
independent of the unknown coupling $c_{\widetilde{G}\Phi}$ in
Table \ref{GPS}. Here, $\Gamma_{\tilde{G}}^{tot}=\Gamma_{\tilde{G}\rightarrow
PPP}+\Gamma_{\tilde {G}\rightarrow PS}$ is the total decay width. Note that in the eLSM (\ref{intlag}) the decay of the pseudoscalar
glueball into three pions vanishes \cite{Eshraim}:
\begin{equation}
\Gamma_{\tilde{G}\rightarrow\pi\pi\pi}=0\text{ .}%
\end{equation}

\begin{table}
\centering
\caption{Branching ratios for the two- and three-body decay of the
pseudoscalar glueball $\tilde {G}$ into a scalar and a
pseudoscalar meson, where $\sigma_{S}$ is assigned either to
$f_{0}(1710)$ or to $f_{0}(1500)$
(values in parentheses).}
\label{GPS}       
\begin{tabular}{llll}
\hline Quantity & result in GeV & Quantity & result
in GeV
\\\hline $\Gamma_{\tilde{G}\rightarrow
KK_{S}}/\Gamma_{\tilde{G}}^{tot}$ & $0.060$
&$\Gamma_{\tilde{G}\rightarrow KK\eta}/\Gamma_{\tilde{G}}^{tot}$ &
$0.049$\\\hline $\Gamma_{\tilde{G}\rightarrow
a_{0}\pi}/\Gamma_{\tilde{G}}^{tot}$ & $0.083$ &
$\Gamma_{\tilde{G}\rightarrow
KK\eta^{\prime}}/\Gamma_{\tilde{G}}^{tot}$ & $0.019$ \\\hline
$\Gamma_{\tilde{G}\rightarrow\eta\sigma_{N}}/\Gamma_{\tilde{G}}^{tot}$
& $0.0000026$ &
$\Gamma_{\tilde{G}\rightarrow\eta\eta\eta}/\Gamma_{\tilde{G}}^{tot}$
& $0.016$ \\\hline
$\Gamma_{\tilde{G}\rightarrow\eta^{\prime}\sigma_{N}}/\Gamma_{\tilde{G}}%
^{tot}$ & $0.039$ &
$\Gamma_{\tilde{G}\rightarrow\eta\eta\eta^{\prime}}/\Gamma_{\tilde{G}}^{tot}$
& $0.0017$ \\\hline
$\Gamma_{\tilde{G}\rightarrow\eta\sigma_{S}}/\Gamma_{\tilde{G}}^{tot}$
& $0.012$ $(0.015)$ &
$\Gamma_{\tilde{G}\rightarrow\eta\eta^{\prime}\eta^{\prime}}/\Gamma_{\tilde
{G}}^{tot}$ & $0.00013$ \\\hline
$\Gamma_{\tilde{G}\rightarrow\eta^{\prime}\sigma_{S}}/\Gamma_{\tilde{G}}%
^{tot}$ & $0$ $(0.0082)$ &$\Gamma_{\tilde{G}\rightarrow
KK\pi}/\Gamma_{\tilde{G}}^{tot}$ & $0.47$ \\\hline
$\Gamma_{\tilde{G}\rightarrow\eta\pi\pi}/\Gamma_{\tilde{G}}^{tot}$
& $0.16$ &
$\Gamma_{\tilde{G}\rightarrow\eta^{\prime}\pi\pi}/\Gamma_{\tilde{G}}^{tot}$
& $0.095$ \\\hline
\end{tabular}
\end{table}

\section{Phenomenology of charmed mesons}

In this section we enlarge the eLSM (\ref{lag}) by including the charm quark \cite{WFDP, WP, WFP}. We then compute the masses of light and (open and hidden)
charmed mesons, the
weak decay constants, and the decay widths of open charmed mesons.\\
In the extension to the four-flavor case, the multiplet $\Phi$ as seen in Eq.(\ref{phimatex}) is extended as follows:
\begin{equation}
\Phi=\frac{1}{\sqrt{2}}
\left(%
\begin{array}{cccc}
  \frac{(\sigma_{N}+a^0_{0})+i(\eta_N +\pi^0)}{\sqrt{2}} & a^{+}_{0}+i \pi^{+} & K^{*+}_{0}+iK^{+} & D^{*0}_0+iD^0 \\
  a^{-}_{0}+i \pi^{-} & \frac{(\sigma_{N}-a^0_{0})+i(\eta_N -\pi^0)}{\sqrt{2}} & K^{*0}_{0}+iK^{0} & D^{*-}_0+iD^{-} \\
  K^{*-}_{0}+iK^{-} & \overline{K}^{*0}_{0}+i\overline{K}^{0} & \sigma_{S}+i\eta_{S} & D^{*-}_{S0}+iD^{-}_S\\
  \overline{D}^{*0}_0+i\overline{D}^0 & D^{*+}_0+iD^{+} & D^{*+}_{S0}+iD^{+}_S & \chi_{C0}+i\eta_C\\
\end{array}%
\right)\,.\label{Phi}
\end{equation}
 The new (pseudo)scalar charmed mesons appear in the
 fourth line and fourth column. In the
 pseudoscalar sector there are: an open charmed state $D^{0,\pm}$,
 open strange-charmed states $D^\pm_S$, and a hidden charmed ground
 state $\eta_C(1S)$ \cite{WFDP}. In the scalar sector there are open charmed
 $D^{\ast\,0,\pm}_0$ and strange charmed meson $D^{\ast\,\pm}_{S0}$ which are
 assigned to $D^\ast_0(2400)^{0,\pm}$ and $D_{S0}^\ast(2317)^\pm$ \cite{WFDP},
 respectively. Similarly, the new (axial-)vector charmed mesons appear also in the fourth line and fourth column of the left- and right-handed matrices which are described 
 as \cite{WFP}:
\begin{equation}\label{4}
L^\mu=V\mu+A^\mu=\frac{1}{\sqrt{2}}
\left(%
\begin{array}{cccc}
  \frac{\omega_N+\rho^{0}}{\sqrt{2}}+ \frac{f_{1N}+a_1^{0}}{\sqrt{2}} & \rho^{+}+a^{+}_1 & K^{*+}+K^{+}_1 & D^{*0}+D^{0}_1 \\
  \rho^{-}+ a^{-}_1 &  \frac{\omega_N-\rho^{0}}{\sqrt{2}}+ \frac{f_{1N}-a_1^{0}}{\sqrt{2}} & K^{*0}+K^{0}_1 & D^{*-}+D^{-}_1 \\
  K^{*-}+K^{-}_1 & \overline{K}^{*0}+\overline{K}^{0}_1 & \omega_{S}+f_{1S} & D^{*-}_{S}+D^{-}_{S1}\\
  \overline{D}^{*0}+\overline{D}^{0}_1 & D^{*+}+D^{+}_1 & D^{*+}_{S}+D^{+}_{S1} & J/\psi+\chi_{C1}\\
\end{array}%
\right)^\mu,
\end{equation}
$$\\$$
\begin{equation}\label{5}
R^\mu=V\mu-A^\mu=\frac{1}{\sqrt{2}}
\left(%
\begin{array}{cccc}
  \frac{\omega_N+\rho^{0}}{\sqrt{2}}- \frac{f_{1N}+a_1^{0}}{\sqrt{2}} & \rho^{+}-a^{+}_1 & K^{*+}-K^{+}_1 & D^{*0}-D^{0}_1 \\
  \rho^{-}- a^{-}_1 &  \frac{\omega_N-\rho^{0}}{\sqrt{2}}-\frac{f_{1N}-a_1^{0}}{\sqrt{2}} & K^{*0}-K^{0}_1 & D^{*-}-D^{-}_1 \\
  K^{*-}-K^{-}_1 & \overline{K}^{*0}-\overline{K}^{0}_1 & \omega_{S}-f_{1S} & D^{*-}_{S}-D^{-}_{S1}\\
  \overline{D}^{*0}-\overline{D}^{0}_1 & D^{*+}-D^{+}_1 & D^{*+}_{S}-D^{+}_{S1} & J/\psi-\chi_{C1}\\
\end{array}%
\right)^\mu\,.
\end{equation}
In the vector sector, the nonstrange-charmed fields $D^{\ast 0},\,D^{\ast\pm }$ correspond to $\overline{q}q$ resonances $D^{\ast }(2007)^{0}$ and
$D^{\ast }(2010)^{\pm }$, respectively, while the strange-charmed $D_{0}^{\ast\pm}$ is assigned to the resonance $D_{0}^{\ast\pm}$ (with mass $m_{D_{0}^{\ast\pm}}$), and there is the lowest vector charmonium state $J/\psi(1S)$. In the axial-vector sector, the open charmed mesons $D_{1}$ and $D_{S1}$ are assigned to $D_{1}(2420)$ and $D_{S1}(2536)$, respectively, whereas the charm-anticharm state $\chi _{c1}$ corresponds to the $c\overline{c}$ resonance $\chi _{c1}(1P)$ \cite{WFDP}.\\
In order to implement the spontaneous symmetry breaking we shift the charm-anticharm scalar field $\chi_{C0}$ by its vacuum expectation value $\phi_C$ \cite{WFDP, WP, WFP} as 
\begin{equation}
\chi_{C0} \rightarrow \chi_{C0}+\phi_C\,,
\end{equation} 
besides shifting the light scalar-isoscalar fields as seen in Eq. (\ref{shift}). Almost all parameters are determined in the low-energy study of Ref.\cite{dick}. We fix the three new parameters related to the charm sector by a fit including twelve masses of hidden and open charmed mesons with $\chi^2/\rm{d.o.f} \simeq 1$ (for details see Ref.\cite{WFDP}).\\

\subsection{Masses}

In the case of the $U(4)_R \times U(4)_L$ linear sigma model, there are interesting mass differences between vector and axial-vector charmed mesons which depend on the charm condensate $\phi_{C}$, and are independent of the parameters $\varepsilon_{C}$ and $\delta_{C}$ \cite{WFDP}:
\begin{equation}
\,\,\,\,m_{D_{1}}^{2}-m_{D^{\ast}}^{2}=\sqrt{2}\,(g_{1}^{2}-h_{3})\phi
_{N}\,\phi_{C},\,\,\,\,\,\,m_{\chi_{C1}}^{2}-m_{J/\psi}^{2}=2\,(g_{1}%
^{2}-h_{3})\phi_{C}^{2},\text{ }m_{D_{S1}}^{2}-m_{D_{S}^{\ast}}^{2}%
=2\,(g_{1}^{2}-h_{3})\phi_{S}\,\phi_{C}\,. \label{massdiff}%
\end{equation}
The theoretical results are
\[
m_{D_{1}}^{2}-m_{D^{\ast}}^{2}=(1.2\pm0.6)\times10^{6}\text{ MeV}^{2}\text{ ,
}\,\,\,m_{\chi_{C1}}^{2}-m_{J/\psi}^{2}=(1.8\pm1.3)\times10^{6}\text{ MeV}%
^{2}\text{, }\]
\[m_{D_{S1}}^{2}-m_{D_{S}^{\ast}}^{2}=(1.2\pm0.6)\times
10^{6}\text{ MeV}^{2}\,,
\]
while the experimental values are 
\[
m_{D_{1}}^{2}-m_{D^{\ast}}^{2}=1.82\times10^{6}\text{ MeV}^{2}\text{ ,
}\,\,\,\,m_{\chi_{C1}}^{2}-m_{J/\psi}^{2}=2.73\times10^{6}\text{ MeV}%
^{2}\text{ , }\,\,\,\,m_{D_{S1}}^{2}-m_{D_{S}^{\ast}}^{2}=1.97\times
10^{6}\text{ MeV}^{2}\,.
\]
The agreement is fairly good, which shows that our determination of the charm
condensate $\phi_{C}$ is compatible with experiment, although it still has
a large uncertainty.\\
\indent We obtained the light and
charmed meson masses, which are reported in Table \ref{ml} \cite{WP} and Table
\ref{mh} \cite{WFDP}, respectively. Note that the values of the light meson masses are the same in
the case of $N_f=3$ as shown in the Ref.\cite{dick}, which are unaffected by the charm sector, because we used the same parameter values as in the case $N_f=3$. \\

\begin{table}
\centering
\caption{ Masses of light meson.}
\label{ml}       
\begin{tabular}{llll}
\hline Resonance & $J^{P}$ &
theoretical value [MeV] & experimental value [MeV]\\
\hline $\pi$          &  $0^{-}$ & 141   &139.57018 $\pm$ 0.00035\\
\hline $\eta$       &  $0^{-}$ & 509   &547.853 $\pm$ 0.024\\
\hline $\eta'$      &  $0^{-}$ & 962   &957.78 $\pm$ 0.06\\
\hline $K$            &  $0^{-}$ & 485  &493.677 $\pm$ 0.016\\
\hline $a_0$        & $0^{+}$ & 1363  &1474 $\pm$ 19\\
\hline $\sigma_1$   & $0^{+}$ & 1362  &(1200-1500)-i(150-250)\\
\hline $\sigma_2$   & $0^{+}$ & 1531   &1720 $\pm$ 60\\
\hline $K_0^\ast$   & $0^{+}$ & 1449   &1425 $\pm$ 50\\
\hline $\omega_N$   & $1^{-}$ & 783   &782.65 $\pm$ 0.12\\
\hline $\omega_S$   & $1^{-}$ & 975   &1019.46 $\pm$ 0.020\\
\hline $\rho$         & $1^{-}$ & 783   &775.5 $\pm$ 38.8\\
\hline $K^*$        & $1^{-}$ & 885    &891.66 $\pm$ 0.26\\
\hline $f_{1N}$     & $1^{+}$ & 1186  &1281.8 $\pm$ 0.6\\
\hline $a_{1}$      & $1^{+}$ & 1185  &1230 $\pm$ 40\\
\hline $f_{1S}$     & $1^{+}$ & 1372  &1426.4 $\pm$ 0.9\\
\hline $K_1$        & $1^{+}$ & 1281  &1272 $\pm$ 7\\
\hline
\end{tabular}
\end{table}

\begin{table}
\centering
\caption{ Masses of charmed
meson }
\label{mh}       
\begin{tabular}{llll}
\hline Resonance & $J^{P}$ & Our
Value [MeV] & Experimental Value [MeV]\\\hline
$D^{0}$ &
$0^{-}$ & $1981\pm73$ & $1864.86\pm
0.13$\\\hline $D_{S}^{\pm}$ & $0^{-}$ &
$2004\pm74$ & $1968.50\pm 0.32$\\\hline $\eta_{c}(1S)$ &
 $0^{-}$ & $2673\pm118$ & $2983.7\pm0.7$\\\hline
$D_{0}^{\ast}(2400)^{0}$ & $0^{+}$ &
$2414\pm77$ & $2318\pm29$\\\hline $D_{S0}^{\ast}(2317)^{\pm}$ &
 $0^{+}$ & $2467\pm76$ &
$2317.8\pm0.6$\\\hline $\chi_{c0}(1P)$  & $0^{+}$ &
$3144\pm128$ & $3414.75\pm 0.31$\\\hline $D^{\ast}(2007)^{0}$ & $1^{-}$ & $2168\pm70$ &
$2006.99\pm0.15$\\\hline $D_{s}^{\ast}$ &
$1^{-}$ & $2203\pm69$ & $2112.3\pm 0.5$\\\hline $J/\psi(1S)$ & $1^{-}$ & $2947\pm109$ & $3096.916\pm 0.011$\\\hline
$D_{1}(2420)^{0}$ &  $1^{+}$ & $2429\pm63$ &
$2421.4\pm0.6$\\\hline $D_{S1}(2536)^{\pm}$ 
& $1^{+}$ & $2480\pm63$ & $2535.12\pm0.13$\\\hline $\chi_{c1}(1P)$
&  $1^{+}$ & $3239\pm101$ & $3510.66\pm 0.07$\\\hline
\end{tabular}
\end{table}

\subsection{Decays}

The weak-decay constants of the pseudoscalar open charmed mesons $D$ and $D_{S}$
\cite{WFDP, WP} are
$$f_{D}=\frac{\phi_{N}+\sqrt{2}\phi_{C}}{\sqrt{2}Z_{D}}=(254\pm17)\text{
MeV },\,\,\,\,\,\,\,\,\,\,\,
f_{D_{S}}=\frac{\phi_{S}+\phi_{C}}{Z_{D_{S}}}=(261\pm17)\text{
MeV }\,,$$
where the experimental values \cite{PDG} are
$$f_{D}=(206.7\pm8.9)\,\text{MeV},\,\,\,\,\,\,\,\,\,\,\,\,\,\,\,\,\,\,\,\,\,\,\,\,f_{D_{s}}=(260.5\pm5.4)\rm{MeV}\,.$$ 
We have a good agreement for
$f_{D_{s}}$ and a slightly too large theoretical result for
$f_{D}$.

The results of (OZI-dominant) strong decay widths of the open
charmed mesons described by the resonances $D_{0}^*,$ $D^{\ast},$ and
$D_{1}$ are summarized in Table \ref{DWCM}
\cite{WFDP, WFP}.

\begin{table}
\centering
\caption{ Decay widths of open
charmed mesons}
\label{DWCM}       
\begin{tabular}{lll}
\hline Decay Channel & Theoretical result [MeV] &
Experimental result [MeV]\\\hline
$D_{0}^{\ast}(2400)^{0}\rightarrow D\pi$ & $139_{-114}^{+243}$ &
$\Gamma=267\pm 40$\\\hline $D_{0}^{\ast}(2400)^{+}\rightarrow
D\pi$ & $51_{-51}^{+182}$ & $\Gamma=283\pm24\pm 34$\\\hline
$D^{\ast}(2007)^{0}\rightarrow D^{0}\pi^{0}$ & $0.025\pm0.003$ &
seen; $<1.3$\\\hline $D^{\ast}(2007)^{0}\rightarrow D^{+}\pi^{-}$
& $0$ & not seen\\\hline $D^{\ast}(2010)^{+}\rightarrow
D^{+}\pi^{0}$ & $0.018_{-0.003}^{+0.002}$ &
$0.029\pm0.008$\\\hline $D^{\ast}(2010)^{+}\rightarrow
D^{0}\pi^{+}$ & $0.038_{-0.004}^{+0.005}$ &
$0.065\pm0.017$\\\hline $D_{1}(2420)^{0}\rightarrow D^{\ast}\pi$ &
$65_{-37}^{+51}$ & $27.4\pm 2.5$\\\hline
$D_{1}(2420)^{0}\rightarrow D^{0}\pi\pi$ & $0.59\pm0.02$ &
seen\\\hline $D_{1}(2420)^{0}\rightarrow D^{+}\pi^{-}\pi^{0}$ &
$0.21_{-0.015}^{+0.01}$ & seen\\\hline $D_{1}(2420)^{0}\rightarrow
D^{+}\pi^{-}$ & $0$ & not seen\\\hline $D_{1}(2420)^{+}\rightarrow
D^{\ast}\pi$ & $65_{-36}^{+51}$ & $\Gamma=25\pm 6$\\\hline
$D_{1}(2420)^{+}\rightarrow D^{+}\pi\pi$ & $0.56\pm0.02$ &
seen\\\hline $D_{1}(2420)^{+}\rightarrow D^{0}\pi^{0}\pi^{+}$ &
$0.22\pm0.01$ & seen\\\hline $D_{1}(2420)^{+}\rightarrow
D^{0}\pi^{+}$ & $0$ & not seen\\\hline
\end{tabular}
\end{table}

\section{Conclusion}

In this work we have studied the phenomenology of the pseudoscalar glueball and charmed mesons in an effective model called the extended linear sigma model (eLSM) with vector and axial vector mesons. Firstly we have presented a chirally invariant effective Lagrangian describing the interaction of the pseudoscalar glueball with (pseudo)scalar mesons. We have computed within the eLSM the decay widths of the lightest pseudoscalar glueball into three pseudoscalar and a scalar and pseudoscalar meson. The results are presented as branching ratios which can be measured in the upcoming PANDA experiment at the FAIR facility \cite{PANDA}. \\
\indent Furthermore, the outline of the extension of the eLSM from the three-flavor case to the four-flavor case including the charm quark has been presented. Most parameters are determined in the low-energy study for the nonstrange-strange sector \cite{dick}. Three new unknown
parameters have been fixed in a fit to the experimental values (details are presented in Ref.\cite{WFDP}). The weak decay constants of nonstrange charm $D$ and strange charm $D_S$ have been calculated. All meson masses in the eLSM (\ref{lag}) have been computed, with the same results for light mesons as in the strange and
non-strange sector \cite{dick}, and with (open and hidden) 
charmed meson masses being in reasonably good
agreement with experimental data \cite{PDG}.  As a last step, we have evaluated the OZI-dominant decays of open
charmed mesons (Table 3). The results are compatible with the results and the upper bounds listed by the PDG \cite{PDG}.\\

Further applications of the described approach are to calculate the decay widths of hidden charmed mesons into light mesons, scalar glueball, and pseudoscalar glueball. This work is currently in progress.

\section*{Acknowledgments}

The author thanks F. Giacosa and D. H. Rischke for cooperation, and S. Janowski for cooperation in the phenomenology of the pseudoscalar glueball. Financial support from the Deutscher Akademischer Austausch Dienst (DAAD) is acknowledged.

\end{document}